\documentstyle[aps,epsfig,multicol]{revtex}                                    %\documentclass[prl,aps,twocolumn,floats,psfig]{revtex} 

\newcommand{\be}{\begin{equation}}
\newcommand{\ee}{\end{equation}}
\newcommand{\bea}{\begin{eqnarray}}
\newcommand{\eea}{\end{eqnarray}}

\newcommand{\p}{\partial}

\newcommand{\rd}{\mbox{d}}

\newcommand{\lb}{\left[}
\newcommand{\rb}{\right]}
\newcommand{\lp}{\left(}
\newcommand{\rp}{\right)}

\def\breakon{\end{multicols}\widetext\vspace{-.2cm}
\noindent\rule{.48\linewidth}{.3mm}\rule{.3mm}{.3cm}\vspace{.0cm}}
 
\def\breakoff{\vspace{-.2cm}
\noindent
\rule{.52\linewidth}{.0mm}\rule[-.27cm]{.3mm}{.3cm}\rule{.48\linewidth}{.3mm}
\vspace{-.3cm}
\begin{multicols}{2}
\narrowtext}
\begin{document}
\draft
%\twocolumn[\hsize\textwidth\columnwidth\hsize\csname @twocolumnfalse\endcsname  
\title{
Narrow gap Luttinger liquid in Carbon nanotubes
%%% Charge compressibility and Wigner crystal in Carbon nanotubes
%%% Charge compressibility of metallic Carbon nanotubes near half-filling
}

\author{L. S. Levitov$^1$ and A. M. Tsvelik$^2$}
\address{$^1$ Department of  Physics, 
%% Center for Material Sciences \& Engineering, 
Massachusetts Institute of Technology, 77 Massachusetts Ave., Cambridge MA02139\\
$^2$ Department of  Physics, Brookhaven National Laboratory, Upton, NY 11973-5000}
\date{\today} 
\maketitle

\begin{abstract}
  Electron interactions reinforce
minigaps induced in metallic nanotubes by an external field and
turn the gap field dependence into a universal power law.
An exactly solvable Gross-Neveau model with an $SU(4)$ symmetry 
is derived for neutral excitations 
near half-filling. Charge excitations, 
described by a sin-Gordon perturbation of Luttinger 
liquid theory,
are composite solitons formed by the charged and neutral fields 
with two separate length scales.
Charge compressibility 
at finite density, 
evaluated in terms of inter-soliton interaction, exhibits a crossover
from overlapping to non-overlapping soliton state. 
Implications for the Coulomb blockade measurements are discussed.
\vskip2mm  
\end{abstract}

\bigskip 

%\pacs{PACS numbers: 71.10.Pm, 72.80.Sk}

\begin{multicols}{2}                                                           \narrowtext
\sloppy

Electron interactions create a peculiar strongly correlated 1D electron 
system~\cite{Kane'97,Egger'97,Balents'97,Krotov'97,Odintsov'99}
in metallic Cabon nanotubes, 
the thinnest and the cleanest among 
the currently available nanoscale quantum wires. 
Luttinger liquid theory of nanotubes predicts~\cite{Kane'97,Egger'97}
that, since tube diameter is larger than Carbon separation, the 1D electron coupling is mainly accounted for by the long-range electron interaction 
(forward scattering), while the  
exchange and Umpklapp scattering, as well as backscattering, are relatively 
weak~\cite{Odintsov'99}. 
Recent experimental work~\cite{Bockrath'99,Egger'00,Bachtold'01} focused on 
Luttinger liquid effects
in tunneling, observed as characteristic 
power laws in the tunneling current dependence on bias voltage and 
temperature. 

Here we discuss Luttinger liquid effects in nanotubes
with a minigap at the band center 
induced by an external perturbation.
Such a minigap can be opened by parallel magnetic field~\cite{Ajiki'93} 
or by the intrinsic curvature of the tube~\cite{KaneMele'97}.
The field- and curvature-induced 
gaps were observed
experimentally~\cite{exp-B-parallel,exp-curvature}
and found to be in agreement with  
the noninteracting electron model. 
We show that electron interaction enhances the charging gap and makes 
it a power law function of the bare gap. 

This provides a unique 
situation, not available in other quantum wires, when 
%% manifestation of the
Luttinger liquid effects are manifest in thermodynamical properties. 
For instance, in a gapped state induced by magnetic field, 
the bare gap is determined without any fitting parameters
by Aharonov-Bohm flux through tube cross section, while the charging gap 
is directly measurable via Coulomb blockade, a prominent feature 
of transport in nanotubes~\cite{Tans'97,Bockrath'97,Tans'98,Cobden'98,Nygard'00,Liang'02}.
In the strong forward scattering limit, the power law relation 
of the charging gap 
and magnetic field is characterized by a universal exponent $4/5$.
We emphasize that the charging gap
measurement is qualitatively different from the tunneling current
measurement because it can be performed in thermodynamic equilibrium. 

Also, in the presence of interactions the charging gap is much enhanced 
compared to the neutral excitation gap,
while in a noninteracting system the two gaps are precisely equal. 
Large energy separation of the charged and neutral sectors results 
in high symmetry of the states in the neutral sector 
described by multiplets of the group $O(6)\sim SU(4)$
derived from exactly solvable Gross-Neveau model. 
This picture, demonstrated in the situation when the gapped state
is created by external field, is realistic, since
intrinsic interaction-induced gap in nanotubes are believed 
to be extremely small\cite{small-gap}

Electron bands of metallic tubes form
%% are described by 
two pairs of spin-degenerate right and left branches intersecting 
at the band center~\cite{Dresselhaus}. The Hamiltonian 
in the forward scattering approximation~\cite{Kane'97,Egger'97} has the form 
\be\label{eq:H0}
{\cal H}_0 = -i\hbar v\int \sum_{j = 1}^4\psi^+_j\sigma_3\partial_x\psi_j dx 
+ \frac{1}{2}\sum_q\rho_qV(q)\rho_{-q}\ , 
\ee
where $\psi_j(x)$ is a two component wavefunction, 
$\rho(x)=\sum_{j = 1}^4\psi^+_j(x)\psi_j(x)$ is charge density,
and $v$ is Fermi velocity. 
The form of the forward scattering amplitude in Eq.(\ref{eq:H0}) 
depends 
on the electrostatic environment. In a nanotube of radius $r$,
in the absence of 
screening, 
$V(q)=\int 
%% {\textstyle \frac{e^2}{|x|}}
V(x)e^{iqx} dx
= e^2 \ln[(qr)^{-2}\!+\!1]$.
The substrate dielectric constant $\epsilon$ reduces $V(q)$ by a factor 
$2/(\epsilon + 1)$. 

Field-induced gapped state~\cite{Ajiki'93,KaneMele'97}
is described by adding to the Hamiltonian (\ref{eq:H0})
a backscattering term
\be\label{eq:Vext}
{\cal V}_{\rm ext} 
= \Delta_0 \int {\textstyle \sum_{j = 1}^4} \psi^+_j\sigma_1\psi_j dx
\ .
\ee
In the absence of interactions, $V(q)=0$, electron spectrum is
$\epsilon(p)=\pm\lp v^2 p^2+\Delta_0^2\rp^{1/2}$ with the value 
of $\Delta_0$ depending on the backscattering mechanism. The magnetic 
field-induced gap~\cite{Ajiki'93} is linear in the field:
$\Delta_0=\hbar v\phi/r$, where $\phi=\pi r^2 B/\Phi_0$ is  
the flux through tube cross-section scaled by $\Phi_0=hc/e$.
For typical tube radius $r\simeq 0.5\,{\rm nm}$ and 
$B\simeq 10\,{\rm Tesla}$, the gap $\Delta_0\simeq 10\,{\rm meV}$.

%%% The form of $V(q)$ depends on the electrostatic environment. 
%%% In the absence of screening, by regularizing the singularity 
%%% of Coulomb potential at the nanotube diameter $d$, 
%%% the interaction takes the form
%%% %
%%% \be
%%% V_0(q) =e^2\!\!\int \frac{e^{iqx}}{|x|}\lp 1\!-\!e^{-|x|/d}\rp dx= 
%%% e^2 \ln\!\lb(qd)^{-2}\!+\!1\rb
%%% %% \ln\lb\frac{1\!\!+\!(qd)^2}{(qd)^2}\rb
%%% \ee
%%% %
%%% %% where $d$ is the nanotube diameter. 
%%% %% [Here the term $e^{-|x|/d}$ is introduced for regularization.]
%%% %% If the nanotube is 
%%% %% parallel to a metallic gate, $V(q) = V_0(q=1/a)=2e^2\ln a/d$, where 
%%% %% $a$ is the distance to the gate (we assume that $a\gg d$ and $|q|\ll 1/a$). 
%%% For a nanotube 
%%% on a substrate with dielectric constant $\epsilon$ the interaction 
%%% changes to $V(q) = \frac2{\epsilon+1}V_0(q)$.

We bosonize the Hamiltonian ${\cal H} = {\cal H}_0 + {\cal V}_{\rm ext}$
in the standard way, 
using $\psi_j\propto e^{i\sqrt{\pi}\Phi_j}/r^{1/2}$. 
The Gaussian part ${\cal H}_0$ is diagonalized 
using linear combinations of the bosonic fields
$\Phi_j = (e_j^{a}\phi_a)$ with
  \bea
%\Phi_j = (e_j^{a}\phi_a), \nonumber\\
{\bf e}_0 = {\textstyle \frac{1}{2}}(1, 1, 1, 1), ~~{\bf e}_1 = {\textstyle \frac{1}{2}}(1, -1, 1, -1),\nonumber\\
{\bf e}_2 = {\textstyle \frac{1}{2}}(1, -1, -1, 1), ~~{\bf e}_3 = {\textstyle \frac{1}{2}}(1, 1, -1, -1)
  \eea
In this notation the Gaussian part of the Lagrangian 
describes one charged and three (neutral) flavor modes:
%% acquires the following form:
  \bea\label{L0-bos}
&&{\cal L}_0 = \frac{1}{2}{\textstyle \sum_q}\left\{\p_{\tau}\phi_0(q)\p_{\tau}\phi_0(-q) 
+ K_q q^2\phi_0(q)\phi_0(-q)\right\} 
\nonumber\\
&& ~~~~~+ \frac{1}{2}\int \rd x {\textstyle \sum_{a = 1}^3}(\p_{\mu}\phi_a)^2 
,\quad
K_q=1 + \frac{4}{\pi}V(q) 
  \eea
$(\hbar=v=1)$.
%% Eq.(\ref{L0-bos}) describes one charged and three neutral modes.
Bosonizing 
%% the interaction 
${\cal V}_{\rm ext}
=\Delta_0 \int dx \sum_{j = 1}^4(R^+_jL_j + {\rm h.c.})$, 
we have
\be\label{Lext-bos}
{\cal L}_{\lambda}=-2\lambda\int \rd x \lp {\textstyle \sum_{j = 1}^4} \cos\sqrt{4\pi}\Phi_i\rp 
\ee
with $\lambda=\Delta_0/r$. 
The total Lagrangian ${\cal L}={\cal L}_0+{\cal L}_{\lambda}$ 
displays the fundamental $U(1)\times SU(4)$ symmetry 
%% which will
playing a major role in our analysis. 
Forward scattering makes the perturbation (\ref{Lext-bos})
even more relevant. At zero chemical potential 
the coupling $\lambda$ grows under renormalization and opens spectral gaps.  
 
There are several possible regimes in which a nanotube, 
described by Eqs.(\ref{L0-bos}),(\ref{Lext-bos}), may exist. 
First, there is an insulating regime with the density at half-filling, 
where all excitations are gapped. 
Second, there are conducting states which can be realized by applying 
various external fields. 
These fields may close some gaps or even all of them,  provided their magnitudes  exceed certain critical values. For example, by varying chemical potential 
one can close all the gaps and make the perturbation $\lambda$ irrelevant. 
This will lead to a transition into a metallic (Tomonaga-Luttinger liquid) 
regime. Fields breaking the $SU(4)$ symmetry will not affect the charge 
sector and thus leave the system insulating, 
but may close some of the gaps in the flavor sector. 

In a nanotube near half-filling, with no external fields present
except $\lambda$, the perturbation (\ref{Lext-bos}) grows under 
RG as $\lambda(l')=(l'/l)^{(1-\alpha)/4}\lambda(l)$, 
where $\alpha=K^{-1/2}$. By the selfconsistency argument $\lambda(l')=1/l'$,
the charge and flavor gaps are related to the noninteracting 
gap $\Delta_0$ as
%% of the noninteracting problem (\ref{eq:Vext}) as 
%% (We assume {\it strong} Coulomb interaction, $K_q\gg1$, and
%% ignore fluctuations of the charge mode $\phi_0$ in (\ref{L0-bos})).
%
\bea\label{eq:gaps}
\Delta_{\rm fl} \simeq  \Delta_0(D/\Delta_0)^{(1-\alpha)/(5-\alpha)} 
,\quad
\Delta_{\rm ch} \simeq  K^{1/2}\Delta_{\rm fl}
,
%% \quad
%% \alpha=K^{-1/2}
\eea
where $D=\hbar v/r$ is 1D bandwidth.
%% and the prefactor $f(K)$ is $\Delta_0$-independent.
%% and $\Delta_0$ is the gap of the noninteracting problem (\ref{eq:Vext}). 
Since $\Delta_0$ is proportional to external magnetic field, 
Eq.(\ref{eq:gaps}) predicts a power law scaling
of the gaps versus an experimentally controllable parameter.
For high charge stiffness $K_q\gg1$,
the gap scaling exponent is universal, with the value $4/5$.
%% :  $\Delta_{\rm ch}\propto\Delta_0^{4/5}$.

The separation (\ref{eq:gaps}) of the charge 
and flavor sector energy scales enables one to integrate out
the fast mode $\phi_0$ in adiabatic approximation.
This will generate an effective action for the flavor modes 
with SU(4) symmetry. Separating $\phi_0$ in
the Lagrangian (\ref{Lext-bos}) one obtains
\bea\label{Lext-bos1}
&&{\cal L}_{\lambda} = - 4\lambda\int \rd x\lp u_1 \cos\tilde\phi_0 + u_2 \sin\tilde\phi_0\rp \\
&&u_1 = {\textstyle \prod_{a=1}^3}\cos\tilde\phi_a, ~~u_2 = {\textstyle \prod_{a=1}^3}\sin\tilde\phi_a
\eea
where $\tilde\phi_i=\sqrt{\pi}\phi_i$.
Treating the slow fields $u_i$ as adiabatic parameters
and shifting the variable
\bea
\phi_0 \rightarrow \phi_0 + \eta, ~~ \eta = \pi^{-1/2}\tan^{-1}\lp u_2/u_1\rp
\eea
we transform the total Lagrangian  
${\cal L}={\cal L}_0+{\cal L}_{\lambda}$ 
as
\bea\label{action}
{\cal L}={\cal L}_0[\phi_0 + \eta]
%% L = L_0[\phi_0] + \sum_{\omega,q}\{\phi_0\}_q
%% %% \hat l_q
%% (\omega^2 + K_q q^2)\{\eta\}_q + L_0[\eta] 
- \int \rd x M[\phi_a]\cos\tilde\phi_0
%% , ~~
%% M[\phi_a] \equiv 4\Delta_0(u_1^2 + u_2^2)^{1/2}
%% =2\Delta_0\lp 1 + \sum_{a \neq b}\cos(2\tilde\phi_a)\cos(2\tilde\phi_b)\rp^{1/2}
\eea
with
$M[\phi_a] 
%% = 4\lambda(u_1^2 + u_2^2)^{1/2}
=2\lambda\lp 1 + {\textstyle\sum_{a \neq b}}\cos 2\tilde\phi_a\cos 2\tilde\phi_b \rp^{1/2}$
%% equals
%
%% \be
%% 2\lambda\lp 1 + {\textstyle\sum_{a \neq b}}\cos 2\tilde\phi_a\cos 2\tilde\phi_b \rp^{1/2}
%% \ee
%	
From Eq.(\ref{action}) we derive an effective action 
for the flavor sector valid for energies well below
the charge gap (\ref{eq:gaps}) by integrating over the fast 
mode $\phi_0$.
%% %% An  estimate for the value of this cut-off will be given later. 
%% As we shall see, the spectral gaps in the flavor sector 
%% %% acquires spectral gaps. We  assume that the value of these gaps is 
%% are much smaller then the 
%% cut-off $\Delta_{\rm ch}$ leaving enough room for renormalization 
%% processes. 
Let us examine the results of this integration. Writing the first term in (\ref{action}) as
${\cal L}_0[\phi_0] + {\cal L}_0[\eta] + \sum_{\omega,q}\{\phi_0\}_q
(\omega^2 + K_q q^2)\{\eta\}_q$
we note that the $\eta$-dependent terms are
%% the term ${\cal L}_0[\eta]$ is 
strongly irrelevant. For example, ${\cal L}_0[\eta]$ contains squares 
of gradients
%% ; however, already the first power of gradient 
%% consists of series of operators with the minimal scaling dimension 2:
%
\bea
\p_{\mu}\eta = \frac{\p_{\mu}\phi_1\sin 2\tilde\phi_2
\sin 2\tilde\phi_3 + \mbox{permut.}}{1 + \sum_{a \neq b}\cos 2\tilde\phi_a\cos 2\tilde\phi_b}
\eea
consisting of series of operators with the minimal scaling dimension 2.
%% It is easy to see that  the second term in Eq.(\ref{action}) 
%% is even more irrelevant.  
The relevant contribution arises from the last term of 
Eq.(\ref{action}). Integrating it over $\phi_0$ we obtain 
the ground state energy of the sine-Gordon model 
\bea
%% &&E_{0} \simeq 
&& - M^{2/(2 - \alpha/4)}[\phi_a] 
%% \approx - M[\phi_a]
%% \nonumber\\
= \mbox{const} - \lambda{\textstyle \sum_{a \neq b}}\cos 2\tilde\phi_a\cos 2\tilde\phi_b + ...  
\nonumber\\
&&~~~~~~~ = - g{\textstyle \sum_{a \neq b}} :\cos 2\tilde\phi_a\cos 2\tilde\phi_b : + ... 
\eea
where $g$ is a suitably renormalized coupling constant
and dots stand for less relevant operators. The normal ordering is taken with respect to the new cut-off $G$. 

Thus at energies  smaller than the cut-off  we obtain the effective action 
for the flavor modes in the form of the O(6)$\sim$ SU(4) Gross-Neveau model:
%
%% \breakon
\bea\label{eq:GrossNeveau}
&& {\cal L} = \int\rd x\lb\frac{1}{2}{\textstyle \sum_{a = 1}^3}(\p_{\mu}\phi_a)^2 -  
g{\textstyle \sum_{a \neq b}}:\cos 2\tilde\phi_a \cos 2\tilde\phi_b:\rb 
\nonumber\\
&&~~~ = \int \rd x\,\lb\, i\bar\chi_j\gamma_{\mu}\p_{\mu}\chi_j - 
g:(\bar\chi_j\chi_j)(\bar\chi_k\chi_k):\, \rb
\eea
%% \breakoff
%
where $\chi_j\, (j = 1,..., 6)$ are Majorana fermions. The latter model is 
exactly solvable, the spectrum consists of $3 + 6 + 3$ relativistic particles 
with masses $m, \sqrt 2 m, m$ transforming according to different 
representations of the O(6)$\sim$ SU(4) group \cite{ZamZam}. 
Their dispersion law is $E_i(p) = (v^2p^2 + m_i^2)^{1/2}$. 
The Majorana fermions themselves belong to the vector representation 
of the group and carry mass $\sqrt 2 m$. 
%% The velocity $v_f$ and t
The mass gap $m$ depends on the bare parameters and 
can be crudely estimated as follows.
%% is difficult to estimate reliably. 
%% However, some estimates can be made. Namely, t
The cut-off $G$ is of order of 
the charge gap $\Delta_{\rm ch}$; it may be somewhat greater than 
$\Delta_{\rm ch}$
because the charge mode has a large velocity 
$v_c=K^{1/2}v \gg v$ 
and therefore can be considered as fast in a larger region of the momentum space than for $v_c = v$. The value of the coupling constant $g$ 
is 
%% determined by the Coulomb interaction so 
such that the mass 
of the Gross-Neveau model (\ref{eq:GrossNeveau}) coincides with the physical 
spin gap (\ref{eq:gaps}):
\bea
G(\Delta_0)\,e^{- 2\pi/3 g} =\Delta_{\rm fl} = f(K)\Delta_0(D/\Delta_0)^{1/5}
\eea
At present we cannot estimate the function $f(x)$.

As for the charge excitations, they carry exactly the same quantum numbers 
as an electron. 
This happens because whenever a soliton of $\phi_0$ is created, 
it acts as an effective potential for flavor fields forming 
bound states. The flavor fields coupled to a $\phi_0$-soliton give it 
the corresponding quantum numbers. 
Therefore,
charge and flavor cannot be separated from each other in the charge sector.

In the sin-Gordon problem (\ref{L0-bos}),(\ref{Lext-bos})
electron is represented by a soliton of one of the fields 
$\Phi_j$. 
Soliton spatial size can be estimated
from a variational principle. Due to large charged stiffness $K$, 
the elastic energy of the soliton is dominated by the field $\phi_0$. 
According to (\ref{L0-bos}) and (\ref{Lext-bos}),
the soliton energy, estimated from the energy in the space interval $l$
where the field $\Phi_j(x)$ varies, is 
$E({\it l})\simeq \lp K_{q{\it l}=1}{\it l}^{-2} + \lambda' \rp {\it l}$
with renormalized coupling $\lambda'(l)=\lambda(r/l)^{(3+\alpha)/4}$.
Soliton size $w$ can be found by minimizing the energy 
with respect to ${\it l}$. Ignoring the logarithmic $l$-dependence of $K$, 
this gives 
$w\approx \lp K/\lambda'(w) \rp^{1/2}$. 
The soliton energy
$E(w) \simeq (K \lambda'(w))^{1/2}$
coincides with the charge gap $\Delta_{\rm ch}$ in (\ref{eq:gaps}).

A peculiar feature of the composite charge soliton 
is the presence of two different length scales, because
the flavor fields $\phi_{1,2,3}$ vary faster than 
the more stiff charge field $\phi_0$.
Let us consider a variational solution 
of the problem ${\cal L}_0+{\cal L}_{\lambda}$ with one of the 
fields $\Phi_j$ varying between two minima of the energy 
(\ref{Lext-bos}). 
To be specific, we consider a soliton of the field
$\Phi_1=\frac12\lp\phi_0+...+\phi_3\rp$ 
in which $\Phi_1$ changes by $\sqrt{\pi}$, while $\Phi_{1,2,3}$ do not 
change. In this case all fields $\tilde\phi_{0,...,3}$ change by 
$\pi/2$. The length $w$ over which the field $\phi_0$ changes 
is much larger than that for $\phi_{1,2,3}$.
Thus the variational problem for $\tilde\phi_0$ can be treated
in a $\pi/2$ step approximation for $\tilde\phi_{1,2,3}$, as
\be\label{Phi_0-eqn}
{\textstyle\frac12}K(\tilde\phi_{0}')^2=4\pi\lambda' 
\lb 1-{\rm max}\,(\cos\tilde\phi_{0},\sin\tilde\phi_{0})\rb
\ee
(We treat $K$ as a constant.)
The solution of Eq.(\ref{Phi_0-eqn}) with 
$\tilde\phi_{0}=\pi/4$ at $x=0$, and the $\cos$ and $\sin$ terms
contributing separately in the regions $x>0$, $x<0$,
is
\be\label{Phi_0-single}
\tilde\phi_{0}(x)= 
\cases{2\cos^{-1}\tanh\lp u-x/w\rp,\ &$x<0$\cr
\frac{\pi}2-2\cos^{-1}\tanh\lp u+x/w\rp,\ &$x>0$}
\ee
where $w=(K/4\pi\lambda')^{1/2}=\hbar v/\Delta_{\rm ch}$, 
$u=\tanh^{-1}(\cos\pi/8)$. 

\begin{figure}
\centerline{\psfig{file=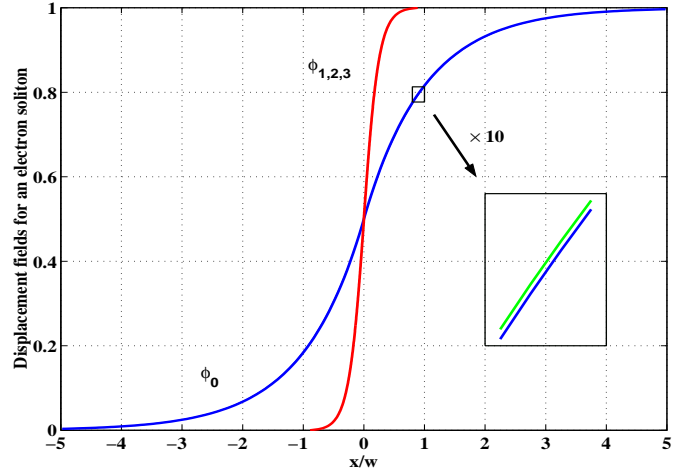,width=3.5in,height=2.5in}}
\vspace{0.1cm}
        \caption[]{
Electron soliton charge and flavor parts
[Eqs.(\ref{Phi_0-single}),(\ref{Phi-neutral}),(\ref{Phi_0-approx})] scaled by $\pi/2$, 
with the stiffness $K=16$. 
The scales for $\tilde\phi_{0}(x)$ and $\tilde\phi_{1,2,3}(x)$
differ by $K^{1/2}$.
The approximate and exact solutions 
(\ref{Phi_0-approx}), (\ref{Phi_0-single})
agree to $0.2\%$.
        }
\label{fig:soliton}
\end{figure}

The problem for the flavor fields 
is simplified because they vary in the region
where $\tilde\phi_{0}\approx \pi/4$.
This gives 
\be\label{Phi-neutral}
{\textstyle \frac32} \tilde\phi'^2
=2^{3/2}\pi\lambda'\lp 1-\cos^3\tilde\phi-\sin^3\tilde\phi\rp
\ee
where $\tilde\phi=\tilde\phi_{1,2,3}$. Integrating Eq.(\ref{Phi-neutral}) 
we obtain the function $\tilde\phi(x)$. 
As illustrated in Fig.\ref{fig:soliton}, the $\tilde\phi_{0}$ step 
is $\sim\sqrt{K}$ times wider than the 
$\tilde\phi_{1,2,3}$ step.
We emphasize that composite solitons  
in which all four fields
$\tilde\phi_{i}$ vary are charge excitations 
of the lowest possible energy.  

%%%%%%%%%%%%%%%%%%%%%%%%%%%%%%%%%%%%%%%%%%%%%%%%%%%%%%%%%%%%%%%%%%%%
%%%%%%%%%%%%%%%%%%%%%%%%%%%%%%%%%%%%%%%%%%%%%%%%%%%%%%%%%%%%%%%%%%%%   

Now we consider multi-soliton solutions and determine charge compressibility 
from inter-soliton interaction. The compressibility
$\chi\equiv (d^2 E_n/d^2n)^{-1}$, with $n$ the electron density, 
is directly related to the capacitance $C=dn/dV_g$ and the
charging spectrum measured in a Coulomb blockade experiment. 
The charging spectrum peak spacing 
is $\delta_n=E_{n+1}-2E_{n}+E_{n-1}\approx \chi^{-1}$. 
We consider densitites smaller than $l_{s}^{-1}$ for which the charge fields $\phi_0$ of 
different solitons may overlap, while the neutral cores with steps of 
$\phi_{1,2,3}$ are isolated.
The effective energy of the 
field $\phi_0$ is obtained by minimizing 
(\ref{Lext-bos1}) with respect to $\tilde\phi_{1,2,3}$
at each value of $\tilde\phi_{0}$, which gives
$-4\lambda'\, {\rm max}_m\cos\lp\tilde\phi_{0}+\frac{\pi}2 m\rp$. 
Switching between different branches occurs at 
$\tilde\phi_0=\frac{\pi}4+\frac{\pi}2 m$.
Approximating 
$\cos$ by a parabola near each maximum 
one obtains
\be\label{Lext-quadratic}
%% {\cal L}_{\lambda} =
U(\tilde\phi_0)=2\lambda'\, {\rm \min}_{m} \lp\tilde\phi_0-\pi m/2 \rp^2
\ee
%
%% arrives at Eq.(\ref{Lext-quadratic}). 
This approximation is extremely accurate, as illustrated in 
Fig.\ref{fig:soliton} by 
comparing Eq.(\ref{Phi_0-single}) with the soliton solution
\be\label{Phi_0-approx}
\tilde\phi_{0}(x)= 
\cases{\frac{\pi}4 e^{x/w},\ &$x<0$\cr
\frac{\pi}2-\frac{\pi}4 e^{-x/w},\ &$x>0$}
\ee
of the variational problem for the total energy 
\be\label{eq:S-energy1}
E[\tilde\phi_{0}]=\int\! dx\! \lp \frac1{2\pi}\p_x\tilde\phi_0 \widehat K \p_x\tilde\phi_0
+ U(\tilde\phi_0)\rp
\ee
with the interaction kernel $\widehat K=\delta(x-x')+\frac4{\pi}V(x-x')$.

In a soliton lattice the field $\tilde\phi_0(x)$ 
is a continuous monotonic function
with smooth $\pi/2$ steps, as in Fig.\ref{fig:soliton}. 
To evaluate the energy (\ref{eq:S-energy1}), 
we introduce a periodic {\it discontinuous} field 
$\varphi=\tilde\phi_0-\pi m/2$ with integer $m$ chosen to
minimize $U(\tilde\phi_0)$.
The energy (\ref{eq:S-energy1}) with
$\tilde\phi_0=\varphi-\frac{\pi}2\sum_m \theta(x-ma)$ takes the form
\be\label{eq:S-energy2}
\int\!\! dx\! \lb\! {\textstyle \frac1{2\pi}}\!\!\lp\p_x\varphi\!+\!{\textstyle \frac{\pi}2} G(x)\rp\widehat K \lp\p_x\varphi\!+\!{\textstyle \frac{\pi}2} G(x)\rp
\!+\! 2\lambda'\varphi^2 \rb
\ee
where $G(x)=\sum_m\delta(x-ma)$. 
The advantage of the form (\ref{eq:S-energy2}) is that 
this {\it quadratic} function can be easily minimized in the Fourier representation. 
We obtain
\be\label{eq:phi-optimal}
\varphi(q)=-iq{\textstyle \frac{\pi}2}G(q)K_q/\lp K_q q^2+4\pi\lambda'\rp
\ee
with $G(q)=2\pi \sum_n\delta(qa-2\pi n)$. 
The energy of (\ref{eq:phi-optimal}) is
\be
E[\varphi]=\sum_q 4\lambda'K_q \lp{\textstyle\frac{\pi}2}G(q)\rp^2 /\lp K_qq^2+4\pi\lambda'\rp
\ee
Using the identity $G^2(q)=L\, G(q)$, with $L$ the system size, the energy 
density can be written as
\be\label{eq:E(n)}
E(n)=\hbar v \frac{\pi n^2}{8} \sum_{m=-\infty}^{\infty} 
\frac{4\pi \lambda' K_{q_m}}{K_{q_m}q_m^2+4\pi\lambda'}
\ee
with $q_m=2\pi m/a$ and soliton density $n=1/a$.
%% number and $q_m=2\pi m/a$.
%% and the soliton spacing $a$ is related with the soliton number $n=L/a$. 

%%%%%%%%%%%%%%%%%%%%%%%%%%%%%%%%%%%%%%%%%%%%%%%%%%%%%%%%%%%%%%%%%%%%
\begin{figure}
\centerline{\psfig{file=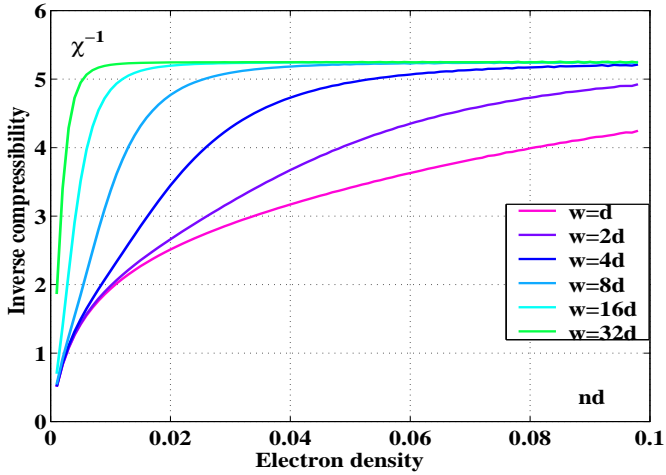,width=3.5in,height=2.5in}}
\vspace{0.1cm}
        \caption[]{
Charge compressibility $\chi=(d^2E(n)/d^2n)^{-1}$ 
scaled by $1/\hbar v$ plotted 
for several soliton widths $w=(4\pi\lambda')^{-1/2}$.
%% $w=(K/4\pi\lambda')^{1/2}$ {\bf better?}.
Eq.(\ref{eq:K-screened}) was used with
the screening length $l_s=0.2\,\mu{\rm m}$,
$v=8\cdot 10^7 {\rm cm/s}$, $\epsilon=12$, tube diameter
$d\equiv 2r=1\,{\rm nm}$.
        }
\label{fig:peak-spacing}
\end{figure}
%%%%%%%%%%%%%%%%%%%%%%%%%%%%%%%%%%%%%%%%%%%%%%%%%%%%%%%%%%%%%%%%%%%%   

Charge compressibility obtained from Eq.(\ref{eq:E(n)}) with 
\be\label{eq:K-screened}
K(q)=1+\frac2{\epsilon+1}\frac{4e^2}{\pi\hbar v}\ln\lb ( q^2+r^{-2})/( q^2+l_s^{-2}) \rb
\ee
that models screening (e.g., by a gate) at distance $l_s\gg r$ 
is plotted in Fig.\ref{fig:peak-spacing}.
At high density $n\gg w^{-1}$, compressibility is density-independent,
$\chi^{-1}_0=\frac{\pi}4\hbar v+\frac{4e^2}{\epsilon+1}\ln (l_s/r)$. 
At lower $n\ll w^{-1}$ it varies 
as $\chi^{-1}=\chi^{-1}_0-\frac{4e^2}{\epsilon+1}\ln (r n)$.  
The crossover can be explained by noting that
neighboring overlapping 
solitons interact via $V(r)\propto r$, while non-overlapping 
solitons interaction is $V(r)\propto 1/r$, because electric field 
is screened and confined to 1D at $r\le w$ and becomes deconfined
at $r\ge w$. 

The crossover density $n\simeq w^{-1}$ is sensitive to the magnetic 
field which controls soliton size $w=\hbar v/\Delta_{\rm ch}$.
Thus the compressibility dependence on soliton size 
(Fig.\ref{fig:peak-spacing}) is manifest in the charging 
spectrum dependence on magnetic field. Along with the power law field 
dependence
(\ref{eq:gaps}) of the gap at half-filling, it represents 
novel 1D electron correlation phenomenon observable 
in a narrow gap state of nanotubes 
in thermodynamic equilibrium.

%% A critical density exists at which
%% a metal-insulator transition takes place. Below this density
%% the soliton interaction changes from $V(r)\propto r$ to $V(r)\propto 1/r$,
%% indicating a deconfinement of electric field in the 1D system.

This work is supported by the MRSEC Program
of the National Science Foundation under Grant No. DMR 98-08941
and by US DOE under contract No. DE-AC02-98CH10886.

\end{multicols}   

\begin{thebibliography}{99}
\bibitem{small-gap}
The leading interaction, forward scattering, 
does not open a gap, while other interactions are 
just marginally relevant in the renormalization group~\cite{Kane'97,Egger'97}
and also small numerically~\cite{Odintsov'99}.
\end{thebibliography}
\end{document}